\documentclass[10pt]{article}

\usepackage{epsfig}
\usepackage{a4wide}
\usepackage{latexsym}
\usepackage{amssymb}
\usepackage{amsmath}

\begin{document}

\def\O{{\cal O}}
\def\N{{\cal N}}
\def\>t{>_{\scriptscriptstyle{\rm T}}}
\def\enu{\alpha_l^{-1}_\nu}
\def\pint{\int{\d^3p\over(2\pi)^3}}   
\def\gint{\int[\D g]\P[g]}
\def\hxi{\hat x_i}
\def\hatx{{\bf \hat x}}
\def\d{{\rm d}}
\def\e{{\bf e}}
\def\x{{\bf x}}
\def\X{{\bf X}}
\def\0x{\x^\smalze}
\def\sperpx{{x_\perp}}
\def\sperpk{{k_\perp}}
\def\sbperpk{{{\bf k}_\perp}}
\def\sbperpx{{{\bf x}_\perp}}
\def\perpx{{x_{\rm S}}}
\def\perpk{{k_{\rm S}}}
\def\bperpk{{{\bf k}_{\rm S}}}
\def\bperpx{{{\bf x}_{\rm S}}}
\def\r{{\bf r}}
\def\q{{\bf q}}
\def\zr{{\bf z}}
\def\R{{\bf R}}
\def\A{{\bf A}}
\def\v{{\bf v}}
\def\u{{\bf u}}
\def\w{{\bf w}}
\def\U{{\bf U}}
\def\cm{{\rm cm}}
\def\l{{\bf l}}
\def\sec{{\rm sec}}
\def\Ckol{C_{Kol}}
\def\flux{\bar\alpha_l^{-1}}
\def\zq{{\zeta_q}}
\def\b{b_{kpq}}
\def\bkab{\bar\kappa_b}
\def\bkas{\bar\kappa_s}
\def\bdu{b^{\scriptscriptstyle (2)}_{kpq}}
\def\z0q{{\zeta^{\scriptscriptstyle{0}}_q}}
\def\smalS{{\scriptscriptstyle {\rm S}}}
\def\smax{{\scriptscriptstyle{\rm MAX}}}
\def\smalel{{\scriptscriptstyle (l)}}
\def\smalze{{\scriptscriptstyle (0)}}
\def\smalun{{\scriptscriptstyle (1)}}
\def\smaldu{{\scriptscriptstyle (2)}}
\def\smaltr{{\scriptscriptstyle (3)}}
\def\smaln{{\scriptscriptstyle (n)}}
\def\smalL{{\scriptscriptstyle{\rm L}}}
\def\smalI{{\scriptscriptstyle {\rm I}}}
\def\smalD{{\scriptscriptstyle{\rm D}}}

\font\brm=cmr10 at 24truept
\font\bfm=cmbx10 at 15truept

\centerline{\brm The behavior of closed}
\vskip 5pt
\centerline{\brm inextensible membranes in}
\vskip 5pt
\centerline{\brm linear and quadratic shear flows}
\vskip 20pt
\centerline{Piero Olla}
\vskip 5pt
\centerline{ISIAtA-CNR}
\centerline{Universit\'a di Lecce}
\centerline{73100 Lecce Italy}
\vskip 20pt

\centerline{\bf Abstract}
\vskip 5pt
The dynamics of a spheroidal vesicle, bounded by an inextensible membrane, is analyzed in
function of the enclosed fluid viscosity, and of the membrane mechanical properties. 
The two situations in which a bending rigidity and a shear elasticity are the potential energy 
source in the membrane have been compared, and the stability properties of quasi-spherical vesicle
shapes discussed. The transition from tank-treading to flipping motion in 
an external shear flow, has been studied in function of the vesicle internal viscosity 
and of the strength of the shear quadratic component. The transverse lift strength has
been calculated in both cases of a tank-treading and of a flipping motion regime.

%\vskip 2cm
%\centerline{{\bf DRAFT} 30/5/99}
%\centerline{\it Submitted to}
%\centerline{\it J. Fluid Mech.}
%\centerline{\it /12/4/98}
\vfill\eject

\centerline{\bf I. Introduction}
\vskip 5pt
Vesicles (closed membranes) can exhibit, in an external fluid flow, a variety of behaviors, 
which are consequence of their ability to change shape in response to the stresses from 
the suspending fluid.
These behaviors are important to understand the rheology of emulsions and other suspensions of
deformable particles, which play an important role especially in the biological realm 
(think of blood).
Depending on the viscosity
of the membrane and of the fluid it contains, a non-spherical vesicle
in a shear flow may either behave as a rigid object, carrying on a kind of flipping motion,
or it could maintain a more or less fixed shape and orientation, while its interior carries
on a circulating ''tank-treading'' motion.

It is well known that deformable objects, e.g. droplets \cite{leal80}, in channel and Couette 
flows are able to migrate transversally to the flow lines and away from the channel walls. 
In small blood vessels, this phenomenon has long been known and carries the name of 
Fahraeus-Lindqwist effect \cite{azelvadre76}. The role of tank-treading motions in the 
transverse migration of vesicles in bounded shear flows has been considered 
in \cite{olla97a,olla97b} focusing on the problem of a fixed shape ellipsoidal cell and to 
the interaction with a wall bounding the fluid. In the case of droplets, in which the interface
dynamics is governed by surface tension, a tank-treading regime, and hence transverse 
migration, occurs irrespective of the droplet fluid viscosity \cite{chan79}. 

A tank-treading
regime and a stationary ellipsoidal shape has been observed by means of direct numerical
simulation \cite{kraus96} also in the case of inextensible closed membranes with a 
finite bending rigidity, confirming previous experimental results presented
\cite{haas97}. The same result has been obtained analytically in the quasi-spherical
regime, including the effect of thermal fluctuations \cite{seifert99}. In all cases, however,
identical values for the viscosity inside and outside the membrane, were considered.
In practical situations, this is not the case. In blood, for instance, the ratio
of hemoglobin and plasma viscosities, inside and outside a red cell, 
is of the order of five, and to this, the cell membrane viscosity must be added. 

Theoretical analysis, given a prescribed ellipsoidal cell shape \cite{keller82}, leads to expect
the transition to flipping motion in a red cell, for a ratio of inner to outer viscosities 
between three and five. In fact, experimental observations tell us that
red cells in-vivo do not stay in a tank-treading motion condition \cite{tran84}. 
It is to be noticed that the transition threshold mentioned before is a decreasing function 
of the cell departure 
from spherical shape, so that, for a fixed viscosity ratio, a cell will always be able to
tank-tread, provided its shape is sufficiently close to spherical.

Given this state of affair, the migration of objects like red cells in small blood vessels must 
be governed 
by a mechanism more complex than the one valid in the case of droplets, which is based essentially
on orientation fixing by tank-treading motions.
Recently, a simplified mechanism for the transverse migration of red cells in small blood vessels
has been discussed in \cite{olla99a}, based on the stabilizing effect that quadratic shear has
on cell orientation and on the contribution to lift from the kind of deformations 
this produces in the cell. Contrary to what happens in the case of a rigid object \cite{olla99b},
this lift acts also in the external portion of the blood vessel, in which flipping motions
prevail. 

It is clear that an explanation of this mechanism in terms of membrane dynamics must take
into account the inextensibility of the membrane itself. This condition prevents
the vesicle from reacting against the excess of stiffness in its structure by simply 
decreasing the amount of non-sphericity of its shape, which would allow maintaining
its ability to 
tank-tread.  This leaves out any model based
on isotropic elastic membranes \cite{barthes80} or on analogy with droplet dynamics, in which the
vesicle rest shape is a sphere. Likewise, any model based on a prescribed vesicle shape,
like the one considered in \cite{keller82,olla97b} is excluded, due to the need to study 
the flow produced deformations.

Helfrich \cite{helfrich73} and Zhong-can \& Helfrich \cite{zhong87,zhong89}, introduced a
vesicle model characterized by a membrane bending rigidity dominated dynamics, and with the
total area conservation constrain imposed through a Lagrange multiplier term added to the 
energy of the system. Similar approaches were used in the context of fluctuating micro-emulsions
in \cite{safran83,milner87}.  More recently, the
local inextensibility constrain was taken into account, in \cite{cai95} introducing a 
finite compressibility in the membrane dynamics, and in \cite{seifert99} by means of a 
local Lagrange multiplier coupled with the metric tensor of the surface element expressed
in spherical coordinates. In the present paper, a different approach will be used, considering
the dynamics of a finite thickness membrane and taking the zero thickness limit at the
end of the calculation; this with the aim of retaining in the analysis, the effect of the
membrane shear elasticity and viscosity. The range of vesicle dimensions and shear strengths
that will be considered is such that creeping flow conditions can be assumed, but, at the
same time, that thermodynamical fluctuations, both at the vesicle and at the membrane 
scale can safely be neglected.

This paper is organized as follows. In section II, the equation for the membrane shape dynamics
will be derived, starting from a Lamb representation for the fluid flow \cite{happel} and 
for the deformation field inside the membrane. In section III the transition from flipping
to tank-treading motion will be analyzed taking into account the effect of a quadratic 
contribution to shear.
In the tank-treading motion regime, the vesicle shape, orientation and transverse drift 
velocity will be given
in function of the viscosity ratios and of the vesicle non-sphericity. In section IV, 
the time dependent deformations to an initial ellipsoidal shape, and their 
contribution to the transverse drift will be calculated in the flipping motion regime.
Section V will be devoted to discussion of the results and conclusions. The appendix is
devoted to discussion of the relative importance of torque and deformation in leading to 
vesicle alignment in the middle of a channel flow.
\vskip 20pt

\centerline{\bf II The role of shear stresses inside the membrane}
\vskip 5pt
Consider a spheroidal vesicle of volume $\frac{4}{3}\pi R^3$ and area $(4\pi+S)R^2$; $S$ is
the dimensionless excess area parametrizing the deviation from spherical shape. Indicate with 
$\delta$ the membrane thickness, which will be sent to zero at the end of the calculation, 
and with ${\boldsymbol{\Delta}}(\x)$ the displacement of the membrane element with respect 
to an initial position on the spherical surface $x=R$. Since attention is concentrated 
here only on the deformations, that part of the
dynamics associated with bulk translations and rotations is eliminated by working in the 
proper reference frame. For simplicity, the particle is supposed to be neutrally buoyant,
so that the reference frame is, to $\O(\Delta/R)$, the one translating with the fluid and 
rotating with angular frequency equal to half the external flow vorticity 
$\omega_0$ at the vesicle position.

%\begin{figure}[hbtp]\centering
%\centerline{
%\psfig{figure=membrfig1.eps,height=7.cm,angle=-90.}
%}
%\caption{Sketch of non-spherical membrane. The tangential components of the displacement
%vector ${\boldsymbol{\Delta}}$ are important, even in the case of zero shear elasticity
%and viscosity in the membrane, because of the inextensibility constrain.
%}
%\end{figure}
It is convenient to work from the start with units such that $R$, $\omega_0$ and 
the suspending fluid dynamical viscosity $\mu$, are all equal to one. The membrane material
is assumed to be isotropic and elastic, so that the stress tensor is in the form:
${\bf s}=-P{\bf 1}+\frac{\kappa_s}{\delta}{\boldsymbol{\sigma}}+
\frac{\lambda_s}{\delta}\dot{\boldsymbol{\sigma}}$ 
(of course, $\dot{\boldsymbol{\sigma}}\equiv
\partial_t{\boldsymbol{\sigma}}$), where $P$ 
is the pressure, ${\boldsymbol{\sigma}}=\nabla{\boldsymbol{\Delta}}+
(\nabla{\boldsymbol{\Delta}})^\dag$ is the strain tensor, and 
$\kappa_s=\frac{\bar\kappa_s}{\mu\omega_0}$ and $\lambda_s=\mu_s/\mu$ are
respectively the dimensionless counterparts of the shear elasticity and viscosity $\bar\kappa_s$ 
and $\mu_s$. (The factors $\delta^{-1}$ indicate that $\kappa$ and $\lambda$ are quantities 
integrated over the membrane thickness). Once the proper boundary conditions for
${\boldsymbol{\Delta}}$ and $\dot{\boldsymbol{\Delta}}$ are set, the equation for the 
elastic balance inside the medium: $\nabla\cdot{\bf s}=0$ can be split into the two 
independent pieces:
$$
\nabla^2{\boldsymbol{\Delta}}+\nabla (\nabla\cdot{\boldsymbol{\Delta}})=
\frac{\delta}{\kappa_s}\nabla P_\kappa;
\qquad
\nabla^2\dot{\boldsymbol{\Delta}}+\nabla (\nabla\cdot\dot{\boldsymbol{\Delta}})=
\frac{\delta}{\lambda_s}\nabla P_\lambda
\eqno(2.1)
$$
with $P_\lambda+P_\kappa=P$. It is expedient to expand the displacement field ${\boldsymbol{\Delta}}$
into vector spherical harmonics:
$$
{\boldsymbol{\Delta}}(\x)=\sum_{lm}[\Delta^{\rm s}_{lm}(x){\bf Y}^{\rm s}_{lm}(\hxi)+
         \Delta^{\rm e}_{lm}(x){\bf Y}^{\rm e}_{lm}(\hxi)+
         \Delta^{\rm m}_{lm}(x){\bf Y}^{\rm m}_{lm}(\hxi)],
\eqno(2.2)
$$
where $\hxi=x_i/x$ are the cosines of $\x$ in an appropriate reference frame and the
superscripts $\{ {\rm sem}\}$, standing for scalar, electric and magnetic, come from
the origin of this basis as a tool in the study of electromagnetic waves (see \cite{landau4})
In terms of standard spherical harmonics, the functions ${\bf Y}^\mu_{lm}$
are defined as
$$
{\bf Y}^{\rm s}_{lm}(\hxi)=\hatx Y_{lm}(\hxi)\qquad
{\bf Y}^{\rm e}_{lm}(\hxi)=\frac{x\nabla Y_{lm}(\hxi)}{\sqrt{l(l+1)}}\qquad
{\bf Y}^{\rm m}_{lm}(\hxi)=\frac{[\x\times\nabla]Y_{lm}(\hxi)}{\sqrt{l(l+1)}},
\eqno(2.3)
$$
where $\hatx=x^{-1}\x$. The radial component $\Delta^{\rm s}_{lm}$ gives therefore the vesicle
shape deviation from spherical. In particular, it is possible to write for the excess area 
\cite{seifert99}:
$$
S=\frac{1}{2}\sum_{lm}(l^2+l-2)|\Delta^{\rm s}_{lm}|^2
\eqno(2.4)
$$
It is easy to show that $\Delta^{\rm s}_{00}=\O(S)$, while $\Delta^{\rm s}_{lm}=
\O(S^\frac{1}{2})$ for $l>0$. Also, $l=1$ terms are associated with translation of the
vesicle centre, so that all sums over spherical harmonics components actually start at
$l=2$. Substituting Eqns. (2.2-3) into Eqn. (2.1) gives for the pressure:
$$
P=\frac{1}{\delta}\sum_{lm}\Big[x^{-4}(x^4\hat\Delta^{\rm s}_{lm})'+
\frac{(x\hat\Delta^{\rm e}_{lm})''}{\sqrt{l(l+1)}}
-\frac{2}{x}\sqrt{l(l+1)}\hat\Delta^{\rm e}_{lm}\Big]Y_{lm}
\eqno(2.5)
$$
where primes indicate derivation with respect to the radial coordinate $x$, and $\hat\Delta\equiv
\kappa_s\Delta+\lambda_s\dot\Delta$. From here, the force densities ${\bf f}^{out}$ and
${\bf f}^{out}$ exerted by the fluids outside  and inside the membrane, on the membrane itself,
are given, to lowest order in $\Delta$, by $\hat\x\cdot{\bf s}^{out}$
and $-\hat\x\cdot{\bf s}^{in}$; using Eqns. (2.2-3) and (2.5):
$$
\hat\x\cdot{\bf s}=\frac{1}{\delta}\sum_{lm}\Big\{\Big[x^4(\hat\Delta^{\rm s}_{lm}/x^4)'-
\frac{(x\hat\Delta^{\rm e}_{lm})''}{\sqrt{l(l+1)}}
+\frac{2}{x}\sqrt{l(l+1)}\hat\Delta^{\rm e}_{lm}\Big]{\bf Y}^{\rm s}_{lm}
$$
$$
+\Big[x(\hat\Delta^{\rm e}_{lm}/x)'+\sqrt{l(l+1)}\hat\Delta^{\rm s}_{lm}/x\Big]
{\bf Y}^{\rm e}_{lm}
+x(\hat\Delta^{\rm m}_{lm}/x)'{\bf Y}^{\rm m}_{lm}\Big\}
\eqno(2.6)
$$
The constitutive equation for the medium must coincide in the zero thickness limit
with the inextensibility condition $\nabla_\perp\cdot{\boldsymbol{\Delta}}(\x)=0$, 
where $\nabla_\perp$ is the angular part of the nabla operator in 
spherical coordinates \cite{note1}; for $\delta$ finite, the constitutive equation  
must be therefore in the form:
$$
\nabla_\perp\cdot{\boldsymbol{\Delta}}(\x)=\gamma^\kappa(\hatx)(x-1);
\qquad
\nabla_\perp\cdot\dot{\boldsymbol{\Delta}}(\x)=\gamma^\mu(\hatx)(x-1).
\eqno(2.7) 
$$
where the function $\gamma$ is still arbitrary.
Putting to system with Eqn. (2.1) and using Eqns. (2.2-3) and (2.5), leads to the following
radial dependence for $\hat{\boldsymbol{\Delta}}$:
$$
\begin{cases}
\hat\Delta^{\rm s}_{lm}=\frac{\sqrt{l(l+1)}}{2}\hat\Delta^{\rm e}_{lm}
+\gamma_{lm}(x-1)\\
\hat\Delta^{\rm e}_{lm}=a_{lm}x^{l+1}+b_{lm}x^{-l}+c_{lm}\Big(x-\frac{4-l(l+1)}{2-l(l+1)}\Big)
+d_{lm}x^\frac{l^2+l-2}{2}\\
\gamma_{lm}=\frac{l^2-l-4}{2\sqrt{l(l+1)}}c_{lm}\\
\hat\Delta^{\rm m}_{lm}=e_{lm}x^l+f_{lm}x^{-l-1}\\
\end{cases}
\eqno(2.8)
$$
The function $\gamma$ in Eqn. (2.7), therefore, is in general non-zero and provides one of 
the six parameters necessary for establishing a force balance at the outer and inner  
membrane surfaces. The total force exerted by the inner and outer fluids on the membrane
is given by, to lowest order in $\delta$:  ${\bf f}^{out}+{\bf f}^{in}=
\delta\partial_x(\hat\x\cdot{\bf s})$. The difference between the force exerted on the outer and 
inner fluids, associated with purely internal shear and compression of the membrane, is given
instead by ${\bf f}^{out}-{\bf f}^{in}=2\hat\x\cdot{\bf s}$. 
Substituting into Eqns. (2.6) and (2.8) and taking the $\delta\to 0$ limit
leads to the following simple expression
for the field $\hat{\boldsymbol{\Delta}}(\x)$:
$$
\begin{cases}
\hat\Delta^{\rm s}_{lm}=\frac{l(l+1)}{8(l^2+l-2)}\Big(2f^{\rm s}_{lm}+
\frac{4f^{\rm e}_{lm}}{\sqrt{l(l+1)}}\Big)\\
\hat\Delta^{\rm e}_{lm}=\frac{2\hat\Delta^{\rm s}_{lm}}{\sqrt{l(l+1)}}\\
\hat\Delta^{\rm m}_{lm}=\frac{2f^{\rm m}_{lm}}{2(l^2+l-2)}\\
\end{cases}
\eqno(2.9)
$$
The first of Eqn. (2.9) is the equation for the membrane shape dynamics. The force terms
at its right hand side are just the opposite of the ones produced on the membrane by the 
fluid stresses and by possible other mechanical effects within the membrane, which are 
not accounted for by Eqn. (2.6).
These last contributions are basically two: the tension force produced by the total area
conservation constrain and the curvature dependent force produced by the membrane bending 
rigidity. The necessity a tension force associated with the total area conservation constrain, 
in addition to the local inextensibility condition produced pressure $[$Eqn. (2.5)$]$, 
is due to the fact that the excess area $S$ depends quadratically on $\Delta$, while the
local area changes are linear. The total area conservation constrain can be imposed
introducing a Lagrange multiplier term proportional to $S$ in the membrane potential energy;
this leads to a contribution to the force 
$$
T\partial S/\partial\Delta^{{\rm s}*}_{lm}=(l^2+l-2)T\Delta^{\rm s}_{lm},
\eqno(2.10)
$$ 
where the Lagrangean multiplier $T$ plays the role of a tension. 
The effect of finite bending rigidity can be accounted for
adding to the membrane energy a curvature term \cite{helfrich73,zhong87,zhong89}, which
leads to the contribution to the force \cite{seifert99}:
$$
-\kappa_b(l^2+l-2)(l+1)\Delta^{\rm s}_{lm},
\eqno(2.11)
$$ 
where $\kappa_b$ is the membrane 
bending rigidity (In terms of the dimensional counterpart $\bar\kappa_b$:
$\kappa_b=\frac{\bar\kappa_b R^2}{\mu\omega_0}$).
Of course, a dissipation term due to bending could easily be added
by considering an analogous term with $\lambda_b\dot\Delta$ in place of $\kappa_b\Delta$,
where $\lambda_b=\mu_b/\mu$ is the ratio of the bending viscosity $\mu_b$ with respect
to the viscosity of the external fluid.

The boundary conditions for the velocity on the membrane are obtained combining the
second of Eqn. (2.9) with the requirement that the velocity on the inner and outer 
membrane surfaces coincide in the limit $\delta\to 0$. 
Indicating with $\bar\U$, $\U$ and $\hat\U$, respectively, the value calculated at $x=R$ 
of the external 
velocity field, of the perturbation induced outside, and of the one induced inside
the membrane, the boundary conditions will be:
$$
\begin{array}{ll}
\hat U^{\rm s}_{lm}=\dot\Delta^{\rm s}_{lm};\quad 
&U^{\rm s}_{lm}=\dot\Delta^{\rm s}_{lm}-\bar U^{\rm s}_{lm}\\
\hat U^{\rm e}_{lm}=\frac{2\dot\Delta^{\rm s}_{lm}}{\sqrt{l(l+1)}};\quad
&U^{\rm e}_{lm}=\frac{2}{\sqrt{l(l+1)}}(\bar U^{\rm s}_{lm}+\dot\Delta^{\rm s}_{lm})
-\bar U^{\rm e}_{lm}\\
\end{array}
\eqno(2.12)
$$
It is then possible to write for the force entering the first of Eqn. (2.9):
$$
\begin{cases}
f^{\rm s}_{lm}=-\kappa_b(l^2+l-2)((l+1)+T)\Delta^{\rm s}_{lm}+
g^{\rm s}_{lm}(\U)-h^{\rm s}_{lm}(\bar\U)+\hat\lambda h^{\rm s}_{lm}(\hat\U)\\
f^{\rm e}_{lm}=
g^{\rm e}_{lm}(\U)-h^{\rm e}_{lm}(\bar\U)+\hat\lambda h^{\rm e}_{lm}(\hat\U)\\
\end{cases}
\eqno(2.13)
$$
where $\hat\lambda=\hat\mu/\mu$ is the inner to outer fluid viscosity ratio,
and $g$ and $h$ are the force densities per unit viscosity, on a spherical surface
bounding a fluid flow from the inside and the outside, calculated respectively in
the case of a velocity field decaying at infinity and at zero.
These can be obtained starting from a Lamb representation for the velocity
field \cite{happel}. In explicit form \cite{olla99b}, for $U(x\to\infty)=0$: 
$$
\begin{cases}
g^{\rm s}_{lm}=
-\frac{2l^2+3l+4}{l+1}U^{\rm s}_{lm}+3\sqrt{\frac{l}{l+1}}U^{\rm e}_{lm}\\
g^{\rm e}_{lm}=
3\sqrt{\frac{l}{l+1}}U^{\rm s}_{lm}-(2l+1)U^{\rm e}_{lm}\\
\end{cases}
\eqno(2.14a)
$$
while, for $U(x\to 0)=0$:
$$
\begin{cases}
h^{\rm s}_{lm}=
-\frac{2l^2+l+3}{l}U^{\rm s}_{lm}+3\sqrt{\frac{l+1}{l}}U^{\rm e}_{lm}\\
h^{\rm e}_{lm}=
3\sqrt{\frac{l+1}{l}}U^{\rm s}_{lm}-(2l+1)U^{\rm e}_{lm}\\
\end{cases}
\eqno(5.14b)
$$
The equation for the membrane shape dynamics can now be obtained substituting 
Eqns. (2.10-14) into the first of Eqn. (2.9), and the result is,
indicating $\Delta_{lm}=\Delta^{\rm s}_{lm}$:
$$
\begin{cases}
\dot\Delta_{lm}+A_l\Delta_{lm}=c_{lm}\\
\sum_{lm}d_l|\Delta_{lm}|^2=S
\end{cases}
\eqno(2.15)
$$
where
$$
\begin{array}{ll}
d_l=\frac{1}{2}(l^2+l-2);\qquad
c_{lm}=\frac{1}{\alpha_l}\Big(\frac{4l^3+6l^2-4l-3}{l(l+1)}\bar U^{\rm s}_{lm}
+\frac{2l+1}{\sqrt{l(l+1)}}\bar U^{\rm e}_{lm}\Big);
\\
A_l=a_l+b_lT;\qquad
a_l=\frac{2d_{lm}}{\alpha_l}(\bkab (l+1)+\frac{4\bkas}{l(l+1)});\qquad
b_l=\frac{2d_{lm}}{\alpha_l};
\\
\alpha_l=\frac{2l^3+3l^2+4}{l(l+1)}+\frac{2l^3+3l^2-5}{l(l+1)}\hat\lambda
+\frac{4(l^2+l-2)}{l(l+1)}\lambda_s+\frac{(l^2+l-2)(l+1)}{4}\lambda_b.
\end{array}
\eqno(2.16)
$$
This is a constrained relaxation equation, in which the forcing $c_{lm}$ comes
from the external shear and the relaxation coefficient $A_l=a_l+b_lT$ 
from the membrane elastic properties and the inextensibility produced tension $T$.
In all this, the viscosity dependent coefficient $\alpha_l$ plays the
role of a mass matrix setting the time scale for the membrane response to 
internal and external forces. Notice that there is an ordering between 
the contribution to viscous and elastic forces coming from different $l$, in that
terms coming from shear are important only at low $l$, while the tension 
and the bending rigidity dominate at large $l$. It is to be noticed that the
fact that (2.15) is a first order differential equation comes from 
the proportionality relation existing between electric and scalar components,
in the second of Eqn. (2.9), and this is at the end just a consequence of the 
local inextensibility condition given by Eqn. (2.7).
\vskip 20pt

\centerline{\bf III. Membrane shape dynamics in viscous shear flow}
\vskip 5pt
To fix the ideas, imagine that the vesicle is immersed in a channel flow. It is then
easy to show that the velocity field in the reference frame translating with the 
vesicle is:
$$
\bar\v=\e_3(x_2+q^{-1}(\frac{1}{3}-x_2^2)),\qquad q=L_r-L_l
\eqno(3.1)
$$
where, $x_2=-L_l$ and $x_2=L_r$ are the coordinates of the walls. 
Back to dimensional units, the flow vorticity will thus be:
$$
\omega_0=\frac{(L_r-L_l)\bar v_0}{L_0^2}
\eqno(3.2)
$$
with $2L_0=L_r+L_l$ the channel width and $\bar v_0$ the fluid velocity at the channel axis, 
measured in the laboratory reference frame. 
The analysis in the previous section was carried out in the reference frame
of the rotating rigid particle (i.e. of the rigid particle with shape equal, at the
given instant, to that of the vesicle).
In  order to proceed, it is necessary to write the vesicle shape equation in the laboratory 
frame. As previously discussed, the particle reference frame rotates with angular frequency, 
in dimensionless
units: $\frac{1}{2}+\O(\Delta)$. Indicating with superscripts $l$ and $r$ components in the
laboratory and rotating reference frames, one will have the relation:
$$
\Delta^l(x^l_i)=\Delta^r(x^r_i)=\exp(\frac{t}{2}M_1)\Delta^r(x_i^l)
\eqno(3.3)
$$
and similar expressions for $c_{lm}$,
where ${\bf M}=-{\rm i}\x\times\nabla$ is basically the quantum mechanical angular momentum 
operator and
$x_2^r=x_2^l\cos(t/2)+x_3^l\sin(t/2)$, $x_3^r=x_2^l\sin(t/2)+x_3^l\cos(t/2)$.
Substituting into Eqn. (2.15) will lead then to the following equation in the laboratory
frame:
$$
\begin{cases}
\dot\Delta_{lm}+A_l\Delta_{lm}+\sum_{m'}\Omega_{mm'}^\smalel\Delta_{lm'}=c_{lm}\\
\sum_{lm}d_l|\Delta_{lm}|^2=S
\end{cases}
\eqno(3.4)
$$
where 
$\Omega_{mm'}^\smalel=-\frac{1}{2}\langle lm|M_1|lm'\rangle$; in explicit form \cite{landau3}:
$$
\langle lm|M_1|l,m+1\rangle=\langle l,m-1|M_1|lm\rangle=-\frac{\rm i}{2}\sqrt{(l-m)(l+1+m)}
\eqno(3.5)
$$
Thus, $\Omega_{mm'}^\smalel$ gives nothing else than the variation rate of the components
$\Delta_{lm}$ due to solid rotation at frequency $1/2$.
The tension $T$ is obtained multiplying the first of Eqn. (3.2) by  $d_l\Delta^*_{lm}$,
summing over $lm$ and using the constrain $\dot S=0$; the result is:
$$
T=\frac{\sum_{lm}\alpha_lb_l(c_{lm}\Delta^*_{lm}-a_l|\Delta_{lm}|^2)}{\sum_{lm}\alpha_lb_l^2
|\Delta_{lm}|^2}
\eqno(3.6)
$$
It is important to notice that, in the absence of an external flow, the tension is negative. 
Now, in order for a stable shape configuration to be possible, it is necessary that the 
coefficients $A_l$ in Eqn. (2.15) be positive. This imposes the condition on the
relative importance of membrane elasticity and of the tension, that
the ratio $a_l/b_l$, varying $l$, must be bounded below by some positive constant.
Looking at Eqn. (2.15), one realizes therefore that a membrane dynamics, based purely 
on shear elasticity, is unstable in the quasi-spherical regime, and would require a
non-perturbative treatment. 
In the case the elasticity due to bending dominates the one due to shear (it is enough
that $\kappa_b>\kappa_s/3$) one finds instead that the 
configuration in which all the excess area is concentrated in modes $l=2$ is linearly 
stable. 

\vskip 5pt
\noindent{\bf A. Flipping and tank-treading motion regime in linear shear flow}
\vskip 5pt
Away from the channel center, the linear part of the shear is
dominant. From the analysis carried on in \cite{keller82}, it is known that, in the case
of a prescribed ellipsoidal shape, a vesicle undergoes a transition from tank-treading to flipping
motion, which depends on the inner to outer fluid viscosity ratio and on the 
cell non-sphericity. It is interesting to consider the same problem taking fully into 
account the membrane shape dynamics.

For the flow given in Eqn. (4.1), the only non-zero components are \cite{olla99b}
$\bar U^{\rm s}_{2,\pm 1}={\rm i}\sqrt{\frac{2\pi}{15}}$ and $\bar U^{\rm e}_{2,\pm 1}
={\rm i}\sqrt{\frac{\pi}{5}}$. This leads to:
$$
c_{lm}={\rm i}\delta_{l2}(\delta_{m1}+\delta_{m,-1})c_2
\quad{\rm where}\quad
c_2=\frac{2}{\alpha_2}\sqrt{\frac{10\pi}{3}}
\eqno(3.7)
$$
%From the form of the $l=2$ spherical harmonics, in particular
%$Y_{2,\pm 1}=\sqrt{\frac{15}{8\pi}}\hat x_3(\mp\hat x_1-{\rm i}\hat x_1)$ and
%$Y_{2,\pm 2}=\frac{1}{4}\sqrt{\frac{15}{2\pi}}(\hat x_1^2-\hat x_2^2\pm 2{\rm i}\hat x_1\hat x_2)$,
%one expects a solution with $\Delta_{2\pm 1}$ imaginary and $\Delta_{2\pm 2}$ real, with 
%$\Delta_{2m}=\Delta_{2,-m}$. 
Tank-treading solutions to Eqn. (3.2) correspond to a constant shape configuration. Imposing the
condition $\dot\Delta_{lm}=0$ in the first of Eqn. (3.2), and using Eqns. (3.3,5) leads then
to the following solution:
$$
\begin{cases}
\Delta_{20}=\frac{\sqrt{3/2}c_2}{1+A_2^2};\\
\Delta_{2,\pm 1}=\frac{{\rm i}A_2c_2}{1+A_2^2};\\
\Delta_{2,\pm 2}=\frac{c_2}{2(1+A_2^2)}
\end{cases}
\eqno(3.8)
$$
and $\Delta_{lm}=0$ for $l\ne 2$.
The coefficient $A_2$ is obtained substituting Eqn. (3.6) into Eqn. (3.4):
$$
A_2= \frac{\alpha_2b_2}{2S}\sum_mc_{2m}\Delta_{2m}^*=\frac{4}{S}\frac{A_2c_2^2}{1+A_2^2}
\quad\Rightarrow\quad
A_2=\Big(\frac{160}{3\alpha_2^2S}-1\Big)^\frac{1}{2}
\eqno(3.9)
$$
From Eqns. (2.15) and (3.5), reality of $A_2$ fixes then the condition on the viscosity
ratios $\hat\lambda$, $\lambda_s$ and $\lambda_b$:
$$
%\lambda+\frac{16}{23}\lambda_s<\frac{32}{23}\Big(\frac{1}{2}\sqrt{\frac{15\pi}{S}}-1\Big)
\alpha_2(\hat\lambda,\lambda_s,\lambda_b)<\alpha_2^\smax=4\sqrt{\frac{10}{3S}}
\eqno(3.10)
$$
and, as in \cite{keller82}, in no way do the membrane elastic properties affect the transition
from flipping to tank-treading motion.

For $\alpha_2>\alpha_2^\smax$, no stationary configurations are possible. One can modify the
analysis leading to Eqn. (3.8) to calculate the rotation velocity of the vesicle at $\theta=0$
(i.e. the stationary orientation at the crossover $\alpha_2=\alpha_2^\smax$; see below).
In this case, it is easy to show that $\Delta_{2,\pm 1}=0$, while $\Delta_{20}$ and 
$\Delta_{2,\pm 2}$ are extremal. This last fact implies that $\dot\Delta_{2m}\ne 0$ 
only for $m=\pm 1$. Repeating the calculation leading to Eqn. (3.6) with 
$c_2+{\rm i}\dot\Delta_{21}$
in place of $c_2$, and imposing the condition $A_2=0$ (from $\Delta_{21}=0$) leads then
to the result: $\dot\Delta_{2,\pm 1}={\rm i}(\frac{S}{4}-c_2)={\rm i}(\frac{S}{4}-
\frac{2}{\alpha_2}\sqrt{\frac{10\pi}{3}})$.

The vesicle shape is described by the equation $X(\x)=1+\sum_m\Delta_{2m}Y_{2m}(\hxi)$. It is
simple at this point to find the semi-axes lengths and orientation of this ellipsoid. If
$\{x'_1x'_2x'_3\}$ indicates the reference frame in which the ellipsoid is diagonal, 
with $x'_1\equiv x_1$ and with $\theta$ the orientation angle of $x'_2$ with respect to 
$x_2$, one can write:
$$
X(\x)=Ax_1^2+Bx_2^2+2Cx_2x_3+Dx_3^2
=A'{x'_1}^2+B'{x_2}^2+2C'x'_2x'_3+D'{x'_3}^2
\eqno(3.11)
$$
where $A'=A$ and:
$$
\begin{cases}
B'=\frac{B+D}{2}+\frac{B-D}{2}\cos 2\theta+C\sin 2\theta\\
C'=2C\cos 2\theta-(B-D)\sin 2\theta\\
D'=\frac{B+D}{2}-\frac{B-D}{2}\cos 2\theta-C\sin 2\theta
\end{cases}
\eqno(3.12)
$$
The orientation angle is then fixed requiring $C'=0$. Using the standard spherical harmonics 
expressions $Y_{20}=\frac{1}{2}\sqrt{\frac{5}{\pi}}\Big(\frac{3}{2}\hat x_3^2-\frac{1}{2}\Big)$,
$Y_{2,\pm 1}=\sqrt{\frac{15}{8\pi}}\hat x_3(\mp\hat x_1-{\rm i}\hat x_1)$ and
$Y_{2,\pm 2}=\frac{1}{4}\sqrt{\frac{15}{2\pi}}(\hat x_1^2-\hat x_2^2\pm 2{\rm i}\hat x_1\hat x_2)$,
together with Eqns. (3.6-7) and (3.9-10) leads then to the result for $\theta$:
$$
\theta=-\frac{1}{2}\tan^{-1}\Big(\frac{160}{3\alpha_2^2S}-1\Big)^\frac{1}{2}
\eqno(3.13)
$$
and for the ellipsoid semi-axes:
$$
\begin{cases}
R_1^2\equiv A'=1\\
R_2^2\equiv B'=1-\frac{3\alpha_2S}{32\pi\cos 2\theta}\\
R_3^2\equiv D'=1+\frac{3\alpha_2S}{32\pi\cos 2\theta}
\end{cases}
\eqno(3.14)
$$
Thus, one has $\theta\in [-\pi/4,0]$, with $\theta=-\pi/4$ corresponding to the limit
$S\to 0$ and $\theta=0$ to the transition from tank-treading to flipping motion, 
for $\alpha_2=\O(S^{-\frac{1}{2}})$. From Eqn. (3.12) and $\theta<0$, this
corresponds to the vesicle major axis (i.e. $R_3$) oriented between the stretching 
direction of the shear flow and that of the velocity $\hat\v$ itself. These results
reproduce those obtained in \cite{keller82}, in the case of a fixed shape 
ellipsoidal vesicle, and in \cite{jeffery22} in that of the rigid ellipsoid.
\vskip 5pt
\noindent{\bf B. The effect of quadratic shear}
\vskip 5pt
The same approach can be used to study the transition to tank-treading motion
induced by quadratic shear. From Eqn. (3.1), the quadratic component of the flow
produces $l=1,2,3$ contributions to $\bar\U$ \cite{olla99b} in the form:
$$
\bar U^{\rm s}_{10}=\frac{4\sqrt{3\pi}}{45q},
\qquad
\bar U^{\rm e}_{10}=-\frac{2\sqrt{6\pi}}{45q},
\qquad
\bar U^{\rm s}_{30}=\frac{2\sqrt{7\pi}}{35q},
\qquad
\bar U^{\rm e}_{30}=\frac{4\sqrt{70\pi}}{105q},
$$
$$
\bar U^{\rm s}_{3,\pm 2}=\frac{1}{q}\sqrt{\frac{2\pi}{105}},
\qquad
\bar U^{\rm e}_{3,\pm 2}=\frac{2\sqrt{70\pi}}{105q}
\qquad
\bar U^{\rm m}_{2,\pm 2}=\mp\frac{2{\rm i}\sqrt{5\pi}}{15}.
\eqno(3.15)
$$
As it is clear from Eqn. (2.9), the magnetic component $\bar U^{\rm m}_{2,\pm 2}$ does not 
contribute to shape dynamics.
Substitution into $c_{lm}$ using Eqn. (2.15) gives, as expected, $c_{10}=0$ and the only
components which survive are:
$$
c_{30}=\frac{c_3}{q},
\quad{\rm and}\quad
c_{3,\pm 2}=\sqrt{\frac{5}{6}}\frac{c_3}{q}
\quad{\rm where}\quad
c_3=\frac{5\sqrt{7\pi}}{6\alpha_3};
\eqno(3.16)
$$
Substituting this result, together with Eqn. (3.5) into the first of Eqn. (3.4), and
imposing again stationarity, leads to the following solution for $\Delta_{3m}$:
%$$
%\begin{cases}
%\hat A_3\Delta_{30}+{\rm i}s\sqrt{3}\Delta_{31}=c_3\\
%{\rm i}s\sqrt{\frac{3}{4}}\Delta_{30}+\hat A_3\Delta_{31}
%+{\rm i}s\sqrt{\frac{5}{8}}=0;\\
%{\rm i}s\sqrt{\frac{5}{8}}\Delta_{31}+\hat A_3\Delta_{32}+
%{\rm i}s\sqrt{\frac{3}{8}}\Delta_{33}=\sqrt{\frac{5}{6}}c_3\\
%{\rm i}s\sqrt{\frac{3}{8}}\Delta_{32}+\hat A_3\Delta_33=0
%\end{cases}
%\eqno(3.17)
%$$
$$
\begin{cases}
\Delta_{30}=-\frac{4A_3c_3}{q\beta}(1-4A_3^2)\\
\Delta_{31}=-\frac{{\rm i}c_3}{\sqrt{3}\, q\beta}(9+44A_3^2)\\
\Delta_{32}=\sqrt{\frac{40}{3}}\frac{A_3c_3}{q\beta}(3+4A_3^2)\\
\Delta_{33}=-\frac{{\rm i}\sqrt{5}\, c_3}{q\beta}(3+4A_3^2)
\end{cases}
\eqno(3.17)
$$
where $\beta=16A_3^4+40A_3^2+9$, 
and again $\Delta_{3m}=\Delta_{3,-m}$. 
The tension $T$ and then the 
coefficients $A_2$ and $A_3$ in Eqns. (3.8) and (3.17) could 
be obtained by substituting the result of these equation into Eqn. (3.6).
The resulting expression for the general situation in which the quadratic
and linear shear components are comparable, corresponding to $q\sim 1$, 
is rather complicated. A simpler expression is obtained for small
$q$.  In this regime, 
most of the excess area is stored in the $\Delta_{l=3}$ components, which 
leads to: $A_3\simeq\frac{5}{S}\sum_mc_{3m}\Delta_{3m}^*$,
and, together with Eqn. (3.17), to the following equation for $A_3$:
$$
A_3^4+10\Big(\frac{1}{4}-\frac{4c_3^2}{3Sq^2}\Big)A_3^2
+\Big(\frac{9}{16}-\frac{5c_3^2}{Sq^2}\Big)=0
\eqno(3.18)
$$
This equation admits real solutions when $\frac{9}{16}<\frac{5c_3^2}{Sq^2}$,
which leads to the condition on $q$:
$$
q<q^\smax=\frac{10}{9\alpha_3}\sqrt{\frac{35\pi}{S}}
%\Big(\frac{85}{12}+\frac{19}{3}\lambda+\frac{29}{3}\lambda_s\Big)^{-1}
\eqno(3.19)
$$
and again, the membrane elastic properties do not play any role.
One thus notices that the condition of $q^\smax$ small means nothing
but the fact that the viscosity of the inner fluid or of the membrane 
are large, so that the vesicle, stays in a state of flipping motion when 
placed in a linear shear flow.
In correspondence to the transition, one has $A_3=0$ and the vesicle shape is given
by:
$$
\begin{cases}
\Delta_{30}=\Delta_{32}=0;\\
\Delta_{31}=-\frac{\rm i}{4}\sqrt{\frac{3S}{5}};\\
\Delta_{33}=-\frac{\rm i\sqrt{S}}{4}
\end{cases}
\eqno(3.20)
$$
meaning that the particle fore-aft line is in the direction of the $x_2$ axis.
In the case $q=0$, i.e. at the middle of a channel flow, one has instead, 
from Eqn. (3.18) $A_3=\sqrt{\frac{80}{3S}}\frac{c_3}{q}$ and then:
$$
\begin{cases}
\Delta_{31}=\Delta_{33}=0\\
\Delta_{30}=\sqrt{\frac{3S}{40}}\\
\Delta_{32}=\sqrt{\frac{S}{16}}
\end{cases}
\eqno(3.21)
$$
corresponding to the particle fore-aft line along the $x_3$ axis, with the 
particle tip in the direction of the flow.
Now, the error in the approximation $\Omega^\smalel_{mm'}\simeq\frac{\rm i}{2}
\langle lm|M_1|lm'\rangle$ $[$see definition after Eqn. (3.4)$]$ is proportional
to the part of the torque on the vesicle coming from the deviation of its shape
from spherical. In the present case, a contribution to the torque comes from 
the interaction between the $\Delta_{l=3}$ components and the quadratic shear
and is directed towards alignment of the particle with the direction of the flow
(i.e. $\Delta_{3,\pm 2}$ and $\Delta_{30}>0$ are the only components remaining).
This torque is $\O(q^{-1}S^\frac{1}{2})$ at the transition and then, for 
$\alpha_3 > S^{-1}$, the
transition to flipping motion occurs for a value of $q$ larger than the
one set by Eqn. (3.19). A quantitative evaluation of this effect is presented in 
the appendix.
\vskip 20pt

\centerline{\bf IV. Transverse drift in the tank-treading and flipping motion regimes}
\vskip 5pt
Once the vesicle shape and membrane motions are known, it is possible to calculate the lift
forces originating from interaction with the shear flow and the walls. The transverse drift
dependence on the particle shape, on the presence of walls bounding the fluid and characteristics 
of the shear flow was analyzed in \cite{olla99b} in the case of rigid particles. The results
presented therein can be summarized as follows.  A
fore-aft symmetric particle will drift transversally due to interaction with the walls mediated
by the linear part of the shear flow. In the case of asymmetric particles, there will also be
one contribution from interaction with the wall, this time mediated by the quadratic part
of the shear, and another from direct interaction with the linear part of the flow, without the
walls playing any role in this case.

The most important difference, in the case of a vesicle, is the possibility of fixed 
orientation provided by tank-treading, but the global picture is not modified. 
There are, however, additional contributions to the lift from the internal motions
of the vesicle. The drift velocity $\v^\smalL$ can be calculated as an adequate combination
of components of the velocity perturbation $\U$ around the vesicle. Given the channel flow
configuration described by Eqns. (3.1-2), one has $\v^\smalL=v^\smalL\e_2$, with:
$$
v^\smalL=\langle L|\U\rangle\Big(\frac{1}{L_l^2}-\frac{1}{L_r^2}\Big)+\langle f|\U\rangle
\eqno(4.1)
$$
where
$$
|L\rangle=\frac{15}{16\pi}[-x_1x_2^2,(1-x_3^2)x_2,-x_2^2x_3]
\quad{\rm and}\quad
|f\rangle=(4\pi)^{-1}\e_2
\eqno(4.2)
$$
and the bra-ket notation $\langle\ |\ \rangle$ stands for the integral over solid angle, of the
arguments scalar product, calculated at $x=1$ \cite{olla99b}. Using Lamb representation, one writes
then Eqn. (4.1) in function of the external flow $\bar U$ and the velocity on the membrane $\hat U$;
working to $\O(\Delta)$:
$$
\langle Z|\U\rangle=\langle Z|\{\hat\U^\smalze-\bar\U\}\rangle
+\langle Z|\hat\U^\smalun\rangle+
\sum_{\mu lm}\langle Z|\Delta |\mu l m\rangle U_{lm}^{\prime\,\mu}
\eqno(4.3)
$$
where $Z=L,f$, $|\mu l m\rangle={\bf Y}_{lm}^\mu$ and:
$$
\begin{cases}
U_{lm}^{\prime\,\rm s}=0\\
U_{lm}^{\prime\,\rm e}=\frac{2l+1}{R}\Big(\frac{3\bar U^{\rm s}_{lm}}{\sqrt{l(l+1)}}-
2\bar U^{\rm e}_{lm}\Big)+2l\hat U^{\rm e}_{lm}\\
U_{lm}^{\prime\,\rm m}=-\frac{2l+1}{R}\bar U^{\rm m}_{lm}+\frac{l+1}{R}\hat U^{\rm m}_{lm}\\
\end{cases}
\eqno(4.4)
$$
The first term to right hand side of Eqn. (4.3) is identically zero, corresponding to the fact
that neutrally buoyant spherical particles do not undergo transverse drift.

\vskip 5pt
\noindent{\bf A. Tank-treading motion regime}
\vskip 5pt
In the case of a 
rigid particle, the velocity term $\hat\U$ gives simply the bulk rotation of the particle
following the flow vorticity, and the only non-zero components are $\hat\Delta^{\rm m}_{1m}$.
In the case of a tank-treading vesicle, this is true only for $\hat\U^\smalze$, while 
$\hat\U^\smalun$
contains contributions which arise from the necessity to maintain constant area locally in the
tank-treading process. The drift $v^\smalL$, therefore, differs from that of an identically shaped
freely rotating rigid particle, by the amount $\langle L|\hat\U^\smalun\rangle(R/L)^2+
\langle f|\hat\U^\smalun\rangle$.
To lowest order, the membrane motion is a constant angular frequency circulation $\tilde\U$, which 
reads, in the case under consideration: 
$$
\tilde\U(\x)=\frac{X(\x)}{2}[\e_1\times{\bf n}]
\eqno(4.5)
$$
where ${\bf n}=\hat\x+\sum_{lm}\Delta_{lm}\nabla Y_{lm}$ is, to $\O(\Delta)$, the outer normal 
to the 
membrane.  The velocity correction $\hat\U^\smalun$ is fixed through the inextensibility condition 
$\nabla_\perp\cdot (\hat\U^\smalun+\tilde\U)=0$. Using Eqn. (4.6) and the fact that $\hat\U$ is 
tangential \cite{note2}:
$$
\U^\smalun=\sum_{lm}
\Big(\e_1\cdot [\x\times\nabla]\Delta\Big)_{lm}
\frac{{\bf Y}^{\rm e}_{lm}}{\sqrt{l(l+1)}}
=\frac{x\nabla [\x\times\nabla \Delta]_1}{l(l+1)}
\eqno(4.6)
$$
plus possible magnetic $l=1$ components, associated with renormalization of the tank-treading 
frequency.
In the case of a linear shear flow, the vesicle has the ellipsoidal shape described by the 
following equation; from Eqns. (3.5-7):
$$
\Delta(\x)=\frac{3\alpha_2S}{32\pi}\Big(\hat x_3^2-\hat x_2^2+
2\Big(\frac{160\pi}{3\alpha^2_2S}-1\Big)^\frac{1}{2}\hat x_2\hat x_3\Big)
\eqno(4.7)
$$
The lift velocity can then be calculated substituting Eqns. (4.2-7) into Eqn. (4.1). The result 
close to the left wall, in  dimensional units, is the following; after some algebra:
$$
v^\smalL=\frac{129}{224\pi}\frac{\alpha_2S\bar v_0R^3}{L_0L_l^2}
\Big(\frac{160\pi}{3\alpha^2_2S}-1\Big)^\frac{1}{2}
\eqno(4.8)
$$
Thus, as expected, the lift goes to zero at the transition to flipping motion.

The quadratic shear dominated regime described by Eqn. (3.19),  corresponds to the region
$L_l\sim L_r$. The width $2y^\smax$ of this region can be obtained from 
Eqns. (3.1) and (3.19): $y^\smax=\frac{5R}{9\alpha_3}\sqrt{\frac{35\pi}{S}}$ and is, as it could
easily be guessed, inversely proportional to the non-sphericity and internal viscosities of 
the cell.  It is easy to see that the contributions to lift coming from the interaction
of the $\Delta_{l=3}$ terms with the linear shear, from that of the 
$\Delta_{l=3}$ terms with the walls, through the quadratic shear, and from that of the 
$\Delta_{l=2}$ terms with the walls, through the linear part of the shear, are respectively
$\O(S^\frac{1}{2})$, $\O(S^\frac{1}{2}L_0^{-3})$ and $\O(S^\frac{1}{2}qL_0^{-2})$
\cite{note3}. 
The first contribution
is therefore dominant. Repeating the same calculations leading to Eqn. (4.8) gives 
now the result:
$v^\smalL=\frac{1}{14}\sqrt{\frac{7}{\pi}}\Delta_{30}
+\frac{1}{42}\sqrt{\frac{210}{\pi}}\Delta_{32}$
Back to dimensional units, from inspection of Eqns. (3.16-18):
$$
v^\smalL=\frac{R^2\bar v_0f(q/q^\smax)}{L^2_0\alpha_3}
\eqno(4.9)
$$
where the function $f$ is obtained using 
in Eqn. (3.17) the value for $A_3$ obtained from solution of Eqn. (3.18). 
The plot of the function $f$ is shown in Fig. 1 below. 
\begin{figure}[hbtp]\centering
\centerline{
\psfig{figure=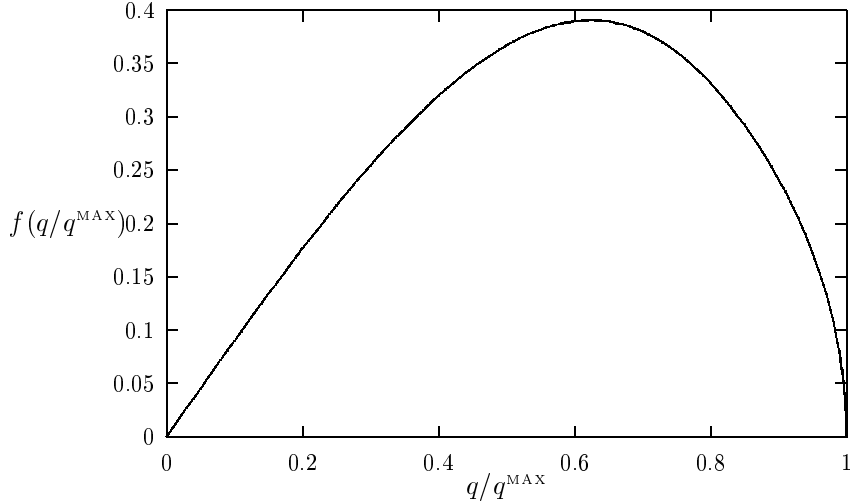,height=7.cm,angle=0.}
}
\caption{Plot of the contribution from quadratic shear to the transverse drift in 
function of the distance of the vesicle from the channel axis. On vertical axis,
$f$ is the transverse drift $v^\smalL$ normalized to 
$\frac{R^2\bar v_0}{L_0^2\alpha_3}$. On horizontal axis, $q/q^\smax$ is just the
vesicle distance to the channel axis, normalized to the half width of the channel
region in which tank-treading motions are possible.
}
\end{figure}
\vskip 5pt
\noindent{\bf B. Flipping motion regime}
\vskip 5pt
Away from the channel axis, the external shear is dominated by its linear component
and the vesicle excess area is expected to be concentrated in the $\Delta_{l=2}$ modes.
In a flipping motion regime, there are two possible sources of lift forces, associated
with the vesicle deformability. One is the possibility of the linear strain, stretching
or compressing the cell in such a way that components $\Delta_{2,\pm 1}$ with a non-zero
average, necessary for a wall induced drift, be present. The other is that the shear 
quadratic piece be able to induce $\Delta_{30}$ and $\Delta_{3,\pm 2}$ components in 
the vesicle shape,
with non-zero average, leading to lift from interaction both with the linear part of the
shear, and with the walls, through the quadratic part of the shear. 

An idea of this effect can be obtained by means of a perturbation theory in 
$\alpha_l^{-1}$: $\Delta=\Delta^\smalze+\Delta^\smalun+...$, with $\Delta_{lm}^\smaln=
\O(\alpha_l^{-n})$. The quadratic part of the shear must still be taken
small enough to be in the flipping motion regime: 
$q\ll q^\smax\sim \alpha_l^{-1} S^{-\frac{1}{2}}$. From inspection of Eqns. (2.15-16) and
(3.4), one finds immediately
$\Omega^\smalel_{mm'}=\O(1)$ and $c_{lm}=\O(\alpha_l^{-1})$. To obtain the order of magnitude
of $A_l$, one needs the tension $T$. As it will be checked a posteriori, to lowest order
in $\alpha_l^{-1}$, only $l=2$ components contribute to $\Delta$. To lowest order one finds
therefore, for the tension: $T=\frac{4}{Sb_2}c_{21}\Delta^{\smalze *}_{21}-\frac{a_2}{b_2}$,
from which one obtains:
$$
A_2=\frac{4}{S}c_{21}\Delta^{\smalze *}_{21}
\quad{\rm and}\quad
A_3=\frac{20\alpha_2}{S\alpha_3}c_{21}\Delta^{\smalze *}_{21}+\frac{10}{\alpha_3}
\Big(\kappa_b-\frac{\kappa_s}{3}\Big)
\eqno(4.10)
$$
Thus, the effect of the membrane elasticity begins to be felt when it comes to extracting 
excess area from the $l=2$ modes and passing it to the higher harmonics. It is interesting
to consider the case in which the viscous and the elastic time scales are comparable, i.e.
$\kappa_b-\kappa_s/3=\O(\alpha_l)$. In this way, $A_2=\O(\alpha_l^{-1})$, while
$A_3\simeq \frac{10}{\alpha_3}(\kappa_b-\frac{\kappa_s}{3})=\O(1)$. Hence, to zeroth order, 
Eqn. (3.4) will read (the convention of summation over repeated indices is adopted from now on):
$$
\begin{cases}
\dot\Delta_{2m}^\smalze+\Omega_{mm'}^\smaldu\Delta_{2m'}^\smalze=0\\
\Delta_{3m}^\smalze=0.
\end{cases}
\eqno(4.11)
$$
while, to $\O(\alpha_l^{-1})$:
$$
\begin{cases}
\dot\Delta_{3m}^\smalun+\Omega^\smaldu_{mm'}\Delta_{2m'}^\smalun=c_{2m}-A_2\Delta_{2m}^\smalze\\
\dot\Delta_{3m}^\smalun+\Omega^\smaltr_{mm'}\Delta_{3m'}^\smalun+A_3\Delta_{3m'}^\smalun=c_{3m}
\end{cases}
\eqno(4.12)
$$
The solution to Eqns. (4.11-12) can be written in the form:
$$
\begin{cases}
\Delta^\smalze_{2m}(t)=d_{mm'}^\smaldu(t)\Delta^\smalze_{2m'}(0)\\
\Delta^\smalun_{2m}(t)=d_{mm'}^\smaldu(t)\Delta^\smalun_{2m'}(0)
+\int_0^t\d\tau d_{mm'}^\smaldu(t-\tau)[c_{2m'}-A_2(\tau)\Delta_{2m'}^\smalze(\tau)]\\
\Delta_{3m}^\smalun(t)=\int_{-\infty}^t\d\tau d_{mm'}^\smaltr(t-\tau)\exp(-A_3(t-\tau))c_{3m'}
\end{cases}
\eqno(4.13)
$$ 
where $d_{mm'}^\smalel(t)$ is the matrix element of the rotation operator: 
$$
d_{mm'}^\smalel(t)=\langle lm|\exp(\frac{t}{2}M_1)|lm'\rangle
=\int\d\Omega Y^*_{lm}(\hat x_i^l)Y_{lm'}(\hat x^r_i).
\eqno(4.14)
$$
In the equation above, as in Eqn. (3.3), superscripts $l$ and $r$ indicate components 
in the laboratory and the
rotating reference frame and ${\bf M}=-{\rm i}\x\times\nabla$. 
(Hence: $\Delta_{lm}^l=d_{mm'}^\smalel\Delta_{lm'}^r$).

From Eqn. (4.13), one obtains the following results. The lowest order
solution $\Delta_{lm}^\smalze$ is exact in the limit of a cell with an infinitely
viscous interior
and, as expected, it gives just a rigid rotation at constant angular velocity of the
initial shape. Thus, if initially only $l=2$ components are present, no higher 
harmonics will be generated at later times. This means that the relaxation term
$A_2$ entering the second of Eqn. (4.13), i.e. the equation for $\Delta^\smalun_{2m}$
will have a purely oscillatory behavior. Then, all factors in the integrand
of that equation will be oscillatory and the corresponding integral will either grow
linearly in time or be itself an oscillating function. Actually, secular behaviors 
could signal the presence of some selection rule for the possible shape of the 
vesicle at order zero. However, it turns out that all factors in the integral are
out of phase and no secular terms are present irrespective of the vesicle shape
(as it might have been expected, perhaps, from the equivalence, from the energetical
point of view of shapes, with different contributions from the different $m$'s but identical
value of $l$ \cite{zhong89}). To be convinced of this fact, it is enough to notice
that $d^\smaldu_{mm'}(\theta)$ is even or odd depending on whether $m+m'$ is itself
even or odd, and that $A_2(t)$ is in the form $c_2(d^\smaldu_{10}(-t)\Delta^l_{20}+
d^\smaldu_{12}(-t)\Delta^l_{22})$. The result is obtained then by direct substitution
into Eqn. (4.13).

This means that to $\O(\alpha_l^{-1})$, $\Delta_{21}$ is zero on the average and no contribution
to drift is present. The only terms which could have a non-zero average are therefore 
$\Delta^\smalun_{3m}$. 
The calculations can be carried out explicitly. From Eqn. (3.15), the only non-zero
components of $c_{3m}$ are those with $m=0,\pm 2$. Hence, the third of Eqn. (4.13) 
becomes:
$$
\Delta_{3m}^\smalun(t)=\frac{5\sqrt{7\pi}}{6q\alpha_3}\int_0^\infty\d\tau 
\exp(-A_3\tau)\Big(d^\smaltr_{m0}(\tau)+\Big(\frac{5}{6}\Big)^\frac{1}{2}
(d^\smaltr_{m,-2}(\tau)+d^\smaltr_{m2}(\tau))\Big)
\eqno(4.15)
$$
From Eqn. (4.9), the contribution to lift from $\Delta_{l=3}$ terms is due only to $m=0,\pm 2$.
After some algebra, one obtains for the rotation matrix coefficients:
$$
\begin{cases}
d^\smaltr_{00}(\tau)+\Big(\frac{5}{6}\Big)^\frac{1}{2}
(d^\smaltr_{0,-2}(\tau)+d^\smaltr_{02}(\tau))=(5\cos^2(\tau/2)-4)\cos(\tau/2)\\
d^\smaltr_{20}(\tau)+\Big(\frac{5}{6}\Big)^\frac{1}{2}
(d^\smaltr_{2,-2}(\tau)+d^\smaltr_{22}(\tau))=
\sqrt{30}(\frac{1}{2}\cos^2(\tau/2)-\frac{1}{3})\cos(\tau/2)
\end{cases}
\eqno(4.16)
$$
Substituting into Eqn. (4.15) and then into Eqn. (4.9), taking for simplicity
$\kappa_s=0$, so that 
$A_3=\frac{10\kappa_b}{\alpha_3}\simeq\frac{5\bar\kappa_bR^2L_0}{\mu\bar v_0\alpha_3}$, gives 
then the final result, in dimensional
units: 
$$
v^\smalL=\frac{20A_3(8A_3^2+3)R^2\bar v_0}{9(4A_3^2+9)(4A_3^2+1)L_0^2\alpha_3}=
\begin{cases}
\frac{100\bar\kappa_bR^4}{27\alpha_3^2\mu L_0};\quad &{\rm for}\quad A_3\to 0\\
\frac{2\mu\bar v_0^2}{9\bar\kappa_bL_0^3};\quad &{\rm for} \quad A_3\to \infty
\end{cases}
\eqno(4.17)
$$
Notice that the $A\to 0$ limit in the first case of Eqn. (4.17) forbids the
unphysical possibility of a finite drift in the absence of a fluid flow, suggested by
vanishing of the $v^\smalL$ dependence on $\bar v_0$. 
As it was to be expected, a transverse drift in the flipping motion regime is hindered 
by a bending rigidity that is too large. Somewhat less intuitive is the fact
that the same thing occurs also in the opposite $\kappa_b/\alpha_3\to 0$ limit.
In this case, $\Delta^\smalun_{3m}$ becomes periodic like $\Delta^\smalun_{2m}$ 
and the average contribution to drift is zero. 
\vskip 5pt

\centerline{\bf V. Conclusion}
\vskip 5pt
The study carried on in this paper has been directed towards three separate goals:
understanding the effect of shear elasticity and viscosity on the dynamics of an
inextensible membrane; obtaining quantitative informations on the tank-treading vs. 
flipping motion behaviors of vesicles in viscous flow; determining what are the 
consequence as regards the possibility of transversal drifts in channel flow 
configurations. 

As regards the first problem, the tension generated from the global area conservation
constrain, as it is clear from Eqns. (2.14-15) and (3.6),
produces a renormalization of the elastic forces acting on the membrane,
which may actually lead to a change of their sign and hence, to the absence
of equilibrium stable configurations. Stability requires that the contributions to 
elastic force $f$ coming from higher harmonics in the deformation $\Delta$, be weighted
by high enough powers of the spherical harmonics index $l$, specifically, 
$f_l\sim l^z\Delta_l$ with $z\ge 2$. It turns out that the case of an 
elastic force purely due to shear leads to unstable membrane behaviors: the excess area is 
transferred to the modes with the highest $l$ available. 
The presence instabilities signaling that a given shape does not correspond any more
to an energy minimum, after an appropriate parameter change, is a well known feature 
of vesicle dynamics \cite{seifert97}. The present situation, however, hints more
towards the presence of limitations in an approach based on a quasi-spherical approximation, 
than towards unability of membrane shear stresses to provide by themselves
an acceptable vesicle dynamics. In the presence of a strong bending rigidity,  
shear elasticity and viscosity provide only a renormalization of the membrane 
elasticity and of the vesicle inner viscosity, concentrated  at small $l$. 
On the other hand, a small shear modulus implies the presence of deformation modes,
proportional to $[\x\times\nabla]Y_{lm}$ $[$see Eqn. (2.9)$]$, which become  easier
to excite than the other modes, which involve changes of shape and therefore membrane
bending. It is then to be remembered that different effective elastic constants for 
deformations at large and small scales, are actually observed in red cells, although 
perhaps in consequence of more complex mechanisms \cite{strey95}.

Concerning the problem of the behavior of a vesicle with a stiff but deformable
interior in a viscous shear flow, the main motivation to the study is to be able to properly 
model red cell dynamics. One of the aspects is understanding in which way such a
system differs from others more or less similar, like for instance droplets, 
elastic micro-capsules, or even rigid non-spherical particles. One crucial aspect
is then the presence of transitions from tank-treading to flipping motion behaviors.

The picture that comes out of the analysis confirms that of Keller and Skalak 
\cite{keller82} in which a fixed ellipsoidal shape was assumed. Basically, 
at least in the quasi-spherical regime considered, the transition to tank-treading
appears to be independent of the membrane elastic properties. One can understand
in a crude way what happens in terms of a balance between external and internal 
viscous stresses. In a tank-treading regime, the internal motions will have a typical
time scale given by the external flow vorticity $\omega_0$, while the part of the
external flow maintaining the fixed orientation will be equal at most to the strain $s$
(it will approach $s$ for a strongly viscous vesicle). 
Given a deviation from spherical shape $\Delta$ and a vesicle size $R$, the balance
between external and internal stresses will have the form:
$\frac{\hat\mu\omega_0\Delta}{R^2}\lesssim\frac{\mu s}{R}$, where $\mu$ and $\hat\mu$
are the external and internal dynamic viscosities. In the case considered in this 
paper, the ratio $s/\omega_0$ can be large if the quadratic part of the shear is 
large, i.e, in terms of the parameter $q$ introduced in Eqn. (3.1), if
$s/\omega_0\sim q^{-1}> 1$.  The condition for tank-treading becomes therefore:
$$
\frac{\hat\mu\Delta}{\mu R}\lesssim q^{-1}
$$
which gives basically the physical content of Eqns. (3.10) and (3.19): tank-treading
is possible either when the cell is not too strongly non-spherical, or when its interior
is not too viscous, or when the quadratic part of the shear dominates.

In real blood cells, the situation is complicated by the presence of a network,
composed mainly of spectrin \cite{bennet93}, providing an elastic skeleton
for the membrane, and it is not clear whether tank-treading could take place without 
actually destroying the network. 
Nonetheless, in the middle part of the channel, it is likely that the cell reaches 
the channel axis before having the time to actually carry out a whole tank-treading cycle,
so that the problem is likely not to be a real issue.  
Clearly, in all this, effects due to the strongly non-spherical shape, in particular,
the torque produced by interaction between odd harmonics in the vesicle shape and quadratic
shear, are likely to become important and should be taken into account in a realistic model
of red cell dynamics.

Knowledge of the membrane dynamics becomes essential when it comes to understanding
cell deformation. It is well known that within a quasi-spherical approximation, no
distinction is made between different shapes involving spherical harmonics $Y_{lm}$
with the same $l$ \cite{zhong89}. Such an approximation has been sufficient, however,
to determine the $l=3$, quadratic shear produced deformation of a vesicle with an 
ellipsoidal rest shape, and to understand its effect in terms of the possibility 
of transverse drifts in a flipping motion regime. The interesting result is that
such deformations have a preferential orientation, and contribute therefore
to transverse drift, only if the membrane elasticity and the vesicle inner 
viscosity have comparable characteristic frequencies, which is the content of
Eqn. (4.17). In other words the membrane
must be neither too stiff nor too ''loose''. The presence of non-zero average $l=3$
shape components corresponds to the picture of 
the vesicle trying to offer to the parabolic flow a parachute kind of shape,
with the concave side towards the coming fluid. In the simplified model presented
in \cite{olla99a}, this effect was parametrized as a contribution to the vesicle
angular velocity, odd in the angle between vesicle symmetry axis and flow direction.
In the present analysis, this contribution appears to be of higher order in the 
shape non-sphericity, and the dominant contribution is the direct interaction
of the average $l=3$ shape component with the external shear linear part.
This effect, however, may play an important role in the case of strongly non-spherical
vesicles like real red cells.

\vskip 10pt
\noindent{\bf Aknowledgements}: I would like to thank Alessandro Tuveri and Dominique Barthes-Biesel
for interesting and helpful conversation. Part of this research was carried on at CRS4 and at the
Laboratoire de Mod\'elisation en M\'ecanique in Jussieu. I would like to thank Gianluigi Zanetti
and Stephane Zaleski for hospitality.
\vskip 20pt

\vskip 5pt
\noindent{\bf Appendix A. Torque on non-spherical vesicle in quadratic shear}
\vskip 5pt
The presence of a torque ${\bf t}=t\e_1$ on the vesicle is equivalent to a change 
in the vorticity of the external flow, in dimensional units, from $\omega_0$ to
$\omega_0+\Delta\omega$, with $\Delta\omega=\frac{t}{8\pi\mu R^2}$. This corresponds
to a redefinition of $q$ in Eqn. (3.1): 
$$
q=L_r-L_l+\Delta q
\eqno({\rm A}1)
$$
where $\Delta q=\Delta\omega/\omega_0$.
In a way perfectly analogous to the case of the drift velocity $[$see Eqns. (4.1-4)$]$, 
it is possible to express this vorticity renormalization using a bra-ket notation 
\cite{olla99b}. Back to dimensionless units:
$$
\Delta q=\langle t|\hat\U^\smalun\rangle+
\sum_{\mu lm}\langle t|\Delta X|\mu l m\rangle U_{lm}^{\prime\,\mu}.
\eqno({\rm A}2)
$$
where $U_{lm}^{\prime\,\mu}$ is given by Eqn. (4.4), and the ket $|t\rangle$ is given by:
$$
|t\rangle =\frac{3}{8\pi}[0,x_3,-x_2].
\eqno({\rm A}3)
$$
which is proportional to ${\bf Y}_{11}^{\rm m}+{\bf Y}_{1,-1}^{\rm m}$. The velocity correction
$\hat\U^\smalun$ is given, as in the calculation of $v^\smalL$ by Eqn. (4.6) and is 
therefore proportional to a combination of electric components ${\bf Y}_{lm}^{\rm e}$,
which are orthogonal to ${\bf Y}_{1,\pm 1}^{\rm m}$. For this reason, $\hat\U^\smalun$
does not contribute in Eqn. (A1). Substituting Eqn. (3.15) into Eqn. (4.4) and carrying 
out the angular integrals in the term $\langle t|\Delta X|\mu l m\rangle$ leads then to the
result, after some algebra:
$$
\Delta q=-\frac{\rm i}{112}\Big(\frac{7}{\pi}\Big)^\frac{1}{2}(23\sqrt{3}\, \Delta_{31}
+15\sqrt{5}\, \Delta_{33})
\eqno({\rm A}4)
$$
At the transition to flipping motion regime, $q=q^\smax$; hence, using the expression 
for $\Delta$ provided by Eqn. (3.20), one obtains:
$$
-\Delta q=\frac{47}{32}\sqrt{\frac{S}{35\pi}},
\eqno({\rm A}5)
$$
that is the amount by which, the width of the 
region at the channel center where tank-treading dominate, gets increased. From comparison
with Eqn. (3.19), this effect will begin to be important when $\alpha_3\sim 70\pi/S$.


\begin{thebibliography}{99}
\bibitem{note1} In terms of coordinates, $\nabla_\perp\cdot{\boldsymbol{\Delta}}=
\partial_\alpha\Delta^\alpha+\Gamma^\alpha_{\alpha\beta}\Delta^\beta$, where
$\Gamma^\alpha_{\beta\gamma}=\e^\alpha\cdot\partial_\beta\e_\gamma$ is  the
Christoffel symbol and 
$\alpha$, $\beta$ and $\gamma$ are the angular coordinates.
\bibitem{note2} The same result could be obtained, expressing $\dot\Delta^{\rm e}_{lm}$
in function of $\dot\Delta^{\rm s}_{lm}$ in the rotating frame, and then transforming back
to the laboratory, after generalizing Eqn. (3.3) to tangential components.
\end{thebibliography}
\end{document}